**The Effect of Cluster Formation on Graphene Mobility**

K. M. McCreary,[1] K. Pi,[1] A. Swartz,[1] Wei Han,[1] W. Bao,[1] C. N. Lau,[1] F. Guinea,[2] M. I. Katsnelson,[3] and R. K. Kawakami[1][‡]

[1]Department of Physics and Astronomy, University of California, Riverside, CA 92521, USA

[2]Instituto de Ciencia de Materiales de Madrid, CSIC, Cantoblanco, E-28049 Madrid, Spain

[3]Institute for Molecules and Materials, Radboud University of Nijmegen, Toernooiveld 1, 6525 ED Nijmegen, The Netherlands

[‡]e-mail: roland.kawakami@ucr.edu

**Abstract:**

We investigate the effect of gold (Au) atoms in the form of both point-like charged impurities and clusters on the transport properties of graphene. Cryogenic deposition (18 K) of Au decreases the mobility and shifts the Dirac point in a manner that is consistent with scattering from point-like charged impurities. Increasing the temperature to room temperature promotes the formation of clusters, which is verified with atomic force microscopy. We find that for a fixed amount of Au impurities, the formation of clusters enhances the mobility and causes the Dirac point to shift back towards zero.

High electronic mobility in graphene is a striking property that is crucial for many of its potential applications.[1, 2] Therefore, understanding the mechanisms that limit the mobility of carriers in graphene is extremely important. It is also of a high conceptual interest, since transport properties of chiral massless fermions are essentially different from those of conventional charge carriers in metals and semiconductors.[3, 4] Currently, typical mobilities are far below the mobility of electrons in graphite and it is generally assumed that extrinsic effects are the cause of this suppression. Charged impurity scattering has received the most attention,[5-9] with the majority of studies modeling the impurities as point-like objects (1/r potential). Recently, theoretical studies considered the physical structure of the charged impurities and found that clusterization of charged impurities can be one of the most important factors influencing their scattering properties.[10]

In this study, we utilize *in situ* transport measurements to investigate the relationship between the clusterization of charged impurities and the electronic transport in graphene. In ultrahigh vacuum (UHV), we deposit gold impurities onto the surface of graphene devices at cryogenic temperatures, which generates *n-type* doping and decreases the mobility in a manner consistent with scattering from point-like charged impurities. As the sample temperature is gradually increased up to room temperature, the increased thermal energy promotes clusterization of the gold impurities, which leads to an increase of the mobility and a decrease of the electronic doping of graphene. The increase in mobility due to cluster formation is consistent with theoretical models[10] and illustrates a mechanism that plays a role in determining the electrical conductivity of graphene.

Single layer graphene is obtained through mechanical exfoliation of Kish graphite onto a SiO$_2$/Si(001) substrate (SiO$_2$ thickness of 300 nm). A suitable single layer flake is identified using optical microscopy and verified through Raman spectroscopy.[11] Standard electron beam lithography techniques are employed to pattern the sample into a four-point probe geometry, with contacting electrodes of Au(100 nm)/Ti(10 nm). Following the completion of all fabrication steps, the graphene device is annealed in an Ar/H atmosphere at 200ºC to remove photoresist and other chemical residues.[12, 13] The sample is then loaded into a UHV system for degassing at 90ºC. Gold impurities are deposited on the graphene surface using a thermal molecular beam epitaxy (MBE) source while the sample temperature is controlled by a variable temperature flow cryostat. Specially designed electrical probes for transport measurements allow all metal deposition and transport measurements to be performed in the same UHV chamber without transferring or changing the sample position throughout the course of the study.

Both graphene and graphite are known to have high surface diffusions, promoting the clustering of materials when deposited at room temperature. Theoretical calculations of adsorption and diffusion energies suggest the deposition of gold onto a graphene surface at cryogenic temperatures will suppress the motion of the gold, minimizing the formation of clusters.[14-19] In this study, the graphene device is cooled to 18 K in UHV prior to measuring the gate dependent conductivity of the clean sample. Gold is then deposited at a rate of $5.0 \times 10^{11}$ atoms/(cm$^2$ sec) in 1 sec intervals for a total of 6 sec while the temperature is held constant at 18 K. The rate of gold deposition is measured using a quartz crystal deposition monitor, and pressures remain below $7 \times 10^{-10}$ torr. The additional presence of gold atoms on the surface is expected to result in charge transfer between gold and graphene as well as affect the graphene mobility without resulting in wave function hybridization.[20-22] Following each deposition, the gate dependent conductivity is measured. The effect of gold deposition on the transport properties is displayed in Figure 1a. Several trends become apparent as the amount of gold increases. The Dirac point ($V_D$) shifts toward more negative voltages, indicating the transfer of electrons from gold to graphene. With increased gold concentration, the width of the minimum conductivity region increases and the conductivity curves also become more linear. In addition, the slope of the linear portion away from the Dirac point decreases, indicating a decrease in electron and hole mobility, since the mobilities ($\mu_{e,h}$) are determined using the relation $\mu_{e,h} = |\Delta\sigma/e\Delta n|$,[5,7] where the carrier concentration, $n$, is directly related to the gate voltage, $V_g$, by $n=-\alpha(V_g-V_D)$ with $\alpha=7.2\times10^{10}$ V$^{-1}$cm$^{-2}$. The average mobility values, $\mu = (\mu_e + \mu_h)/2$, are shown for the clean sample and the 6 gold depositions in Figure 1b and exhibit a drastic decrease upon the first deposition of gold, followed by a gradual decrease with additional gold.

To determine whether the features described are generated by scattering from point-like charged impurities[5,7] or clusters,[6] it is necessary to investigate the relationship between Dirac point shift ($\Delta V_D = V_D - V_{D,initial}$) and impurity concentration.[5,8] Increasing the impurity concentration (in either adatom or cluster form) will result in more scattering centers and a decreased mobility. The presence of point-like charged impurities affects the mobility through the relationship $\mu n_{imp}=5\times10^{15}$ V$^{-1}$s$^{-1}$.[5,8] Using this relation and the mobility value for the clean graphene, $\mu_0=8390$ cm$^2$/Vs, we find an impurity concentration of $6.0\times10^{11}$ cm$^{-2}$, similar to values found for other high quality clean samples.[23] The quantity $1/\mu - 1/\mu_0$ is therefore proportional to the impurity concentration induced by gold alone. Using the final mobility value of 1270 cm$^2$/Vs after 6 sec of gold deposition, it is calculated that the gold induces $3.3\times10^{12}$ cm$^{-2}$ impurities/cm$^2$, consistent with the value of $3.0\times10^{12}$ atoms/cm$^2$ calculated from the measured deposition rate of gold (previous theoretical work[5, 8] and calculations here assume a charge transfer of one electron per impurity adatom). The additional impurities will also result in a shift of the Dirac point through the power law relationship, $-\Delta V_D \sim (1/\mu - 1/\mu_0)^b$, with values of $b$=1.2-1.3 for point-like scattering (1/$r$ Coulomb potential),[5, 8] while scattering from clusters of material results in $b < 1.0$.[6] Figure 1c shows the data of Dirac point shift vs. $1/\mu - 1/\mu_0$ as well as the best fit line, having a coefficient $b$=1.3 (solid line). For comparison, the dashed line represents a power law with $b$=1. The experimentally measured coefficient, $b$, along with the similarity between measured and calculated impurity concentration indicate that gold deposited at low temperature behaves as point-like charged impurities.

Following the deposition of gold at cryogenic temperatures, the transport properties are monitored as the temperature is increased by discrete amounts until reaching room temperature. Figure 2a shows an increase in the mobility as the temperature is increased, and Figure 2b shows the corresponding change in the Dirac point. At each temperature value, the temperature is stabilized and the gate dependent conductivity is measured to determine the mobility and Dirac point voltage. To test for dynamics, the temperature is held fixed for at least 35 minutes while the transport measurement is repeated every ~8 minutes. In the range between 18 K and 210 K, we observe no time dependence of the mobility or Dirac point. Although, at higher temperatures, the mobility and Dirac point do exhibit a slow dynamics. In Figures 2a and 2b, the multiple data points at 240 K, 270

K, and near RT (292-298 K) represent the time evolution over 42 minutes, 42 minutes, and 600 minutes, respectively. Figures 2c and 2d show the explicit time dependence of the mobility and Dirac point that continues for over 10 hours at RT.

Compared to the initial mobility value of 1270 cm$^2$/Vs at 18 K, the final room temperature value of 3360 cm$^2$/Vs is nearly three times as large. This change in mobility is equivalent to the removal of $2.4 \times 10^{12}$ impurities/cm$^2$. The moderate temperatures used in this study do not provide nearly enough thermal energy to result in evaporation of the gold. Hence, a more plausible explanation for the observed mobility increase involves the rearranging of impurities to form clusters. Based on theoretical predictions, for a fixed amount of impurities, the formation of large circular clusters will decrease the scattering cross section compared to that of isolated adatoms,[10] which would manifest itself experimentally as an increase in mobility. While the mobility shows a continuous change over the measured temperature range, the Dirac point exhibits minimal change between temperature of 75 K and 210 K. This behavior in Dirac point has interesting implications for the doping due to small clusters and will be discussed in more detail later. While clean graphene is known to have minimal temperature dependence, the effect of temperature on metal doped graphene samples has not yet been investigated. To rule this factor out as the cause of the drastic change in doping and scattering observed, the sample is once again cooled. Figure 2e shows the gate dependent conductivity for (1) the initial state at 18 K immediately following the gold deposition (black curve), (2) after heating to room temperature (red curve), and (3) for the second cool-down to 18 K (blue curve). The minimal change observed between room temperature and 18 K during the second cool down, along with the substantial difference between the two curves measured at 18 K, indicates an irreversible change that is due to cluster formation of the gold impurities.

The time dependent properties of the transport measurement provide insight into the clusterization dynamics. At the lower temperatures (18 K – 210 K), no dynamics are observed within the resolution of the measurement (~8 min.). In this regime, the cluster formation is limited primarily by energetics of surface diffusion (i.e. thermal energy vs. energy barrier heights). The clusterization occurs as the temperature is increased and reaches a stable state at fixed temperature within 8 minutes. On the other hand, at higher temperatures (≥240 K), the observed dynamics over long time scales implies that issues of clusterization dynamics (e.g. probability of cluster collisions, density of clusters, etc.) become more important than energetics.

Atomic force microscopy (AFM) is utilized to characterize the structure of the gold on the graphene surface. As suggested by the drastic changes in transport measurements, the gold has formed clusters of material, clearly identified in Figure 3a. The boundary between the substrate and SLG is identified in the figure. There is no preferential clusterization at this boundary, contrary to what has been observed at elevated temperatures.[16] The circular nature of the clusters allows further comparisons between experimental data and the theoretical predications of Katsnelson *et. al.*[10]. By modeling the behavior of charge carriers scattered from a circularly symmetric potential, it is predicted that the scattering cross section of a large cluster ($k_F R > 1$) is comparable to that of a single isolated impurity. Through further analysis of two 1.5 μm × 1.5 μm area scans, the cluster diameter is found to vary from approximately 40 to 120 nm, as seen in Figure 3b. From this data, we are also able to estimate a cluster density of $2.4 \times 10^9$ cm$^{-2}$.

Assuming a large cluster will have a similar effect on mobility as a point like scatterer, a final mobility value on the order of 8300 cm$^2$/Vs is expected (limited by the initial mobility of the clean device). While the observed recovery is clearly not this drastic, it is evident that the clustering has significantly reduced the scattering compared to the initial state of point like charged impurities. A distinct possibility, to account for the only moderate enhancement of mobility, is that not all the material has formed clusters. Single atoms or small clusters of several atoms are below the resolution of AFM. These isolated impurities would provide additional scattering sites, preventing the mobility from fully recovering, without being detected during the AFM measurement. In any case, the measured transport data clearly shows that for the same number of gold atoms, the effect on mobility is significantly reduced when the impurities are in the form of clusters compared to that of isolated adatoms.

While the mobility changes throughout the measured temperature range, the Dirac point shows little variation at low temperatures (< 210 K) followed by a rapid shift toward more positive values as room temperature is approached. We note that the plateau in the Dirac point data (Figure 2b) does not imply that the cluster formation has stopped between 70 K and 210 K; the increase of the mobility in this temperature range clearly indicates that structure of the gold clusters is changing. Rather, the plateau structure is due to a more complicated relation between the electronic doping by clusters and the measurement of the Dirac point. A qualitative understanding of this behavior may be reached by considering the effect of large clusters in comparison to adatoms on the surface of graphene. In the case of adatoms, the graphene system is relatively homogeneous, with the carriers transferred between metal and graphene not localized to any particular region. These transferred carriers result in a large shift in the Fermi energy, as measured by the Dirac point shift. However, for large clusters, their effect on the graphene is more similar to the effect produced by metallic electrodes.[24] In this case, the graphene can no longer be considered a homogeneous system, and is instead separated into three areas. The majority of charge transferred between the metal and graphene is confined to the region directly below an electrode, resulting in a Fermi level that is pinned and unaffected by the gate voltage. Away from the electrode, the carrier concentration is governed solely by the gate voltage. A third region exists at the edges of the electrodes, where charge diffuses away from the large concentration situated below the electrodes. In this transition region, the carrier concentration is affected by both the metal and the gate voltage. It is the presence of this transition region which results in a shift in the Dirac point. While the continually changing mobility in the collected data indicates cluster formation throughout the temperature range, the behavior of the Dirac point implies a transition between the adatom-like and electrode-like situations described above. At lower temperatures, small clusters and adatoms are unable confine carriers, resulting in a Dirac point that has been shifted greatly in comparison to that of the clean sample. Above a certain threshold ($T$~210 K), the clusters behave similarly to electrodes, with some of the doping electrons confined to the region below the cluster and unaffected by gate voltage. As larger clusters continue to form, the number of confined carriers continues to increase. Under the conditions of fixed amount of gold impurities, this leads to a decrease in the number of carriers in the transition regions, causing the Dirac point to shift back towards zero. We note that the Dirac point behavior is consistent with previous studies of Pt clusters on graphene, where the Dirac point shifts toward zero at higher coverages.[6] Therefore, in addition to the effects discussed above, the role of cluster size on the interfacial dipole strength and the equilibrium distance between the metal and graphene may have to be considered.[22, 25]

While this provides a qualitative explanation, future experiments may be able to provide more insight into this behavior by combining transport measurements with low temperature growth and atomic resolution scanning probe microscopy.

In conclusion, we have shown that gold deposited at low temperature behaves as point-like charged impurities. The shift in Dirac point clearly indicates electrons are transferred from gold to graphene at submonolayer coverages. For a fixed amount of gold impurities, it is discovered that the formation of clusters significantly enhances the mobility and causes the Dirac point to shift back toward zero. The latter is attributed to the increased inhomogeneity associated with large clusters, while the former is qualitatively consistent with the theoretical prediction of reduced scattering by clustered impurities.

We acknowledge stimulating discussions with S.-W. Tsai. KMM, KP, AS, WH, and RKK acknowledge the support of ONR (N00014-09-1-0117), NSF (CAREER DMR-0450037), and NSF (MRSEC DMR-0820414). FG acknowledges support by MICINN (Spain) through grants FIS2008-00124 and CONSOLIDER CSD2007-00010, and by the Comunidad de Madrid, through CITECNOMIK. MIK acknowledges support from FOM (the Netherlands). CNL and WB acknowledge the support of NSF (CAREER DMR-0748910) and DOE/DMEA (HH94003-09-2-0901).

FIGURE CAPTIONS:

Figure 1: (a) The gate dependent conductivity at selected values of gold deposition. (b) The mobility as a function of gold deposition. (c) $-\Delta V_D$ is plotted vs. $1/\mu - 1/\mu_0$. The solid line is the power-law fit to the equation, $-\Delta V_D \sim (1/\mu - 1/\mu_0)^b$, where $b=1.3$. For comparison the dashed line shows a power law with $b=1$.

Figure 2: (a) The mobility and (b) Dirac point shift are displayed as a function of temperature. Above 210 K, the mobility and Dirac point change over time at fixed temperature, indicated by the multiple data points at set temperature values. The time dependence of mobility (c) and Dirac point (d) are displayed for the sample at room temperature over a span of 10 hours. (e) A comparison of the gate dependent conductivity curves measured immediately after gold deposition at 18 K, upon warming to room temperature, and after the second cool down indicate irreversible cluster formation.

Figure 3: (a) The room temperature AFM scan shows clusters of gold on the surface of graphene. (b) A histogram of cluster sizes measured in two 1.5 μm × 1.5 μm area scans.

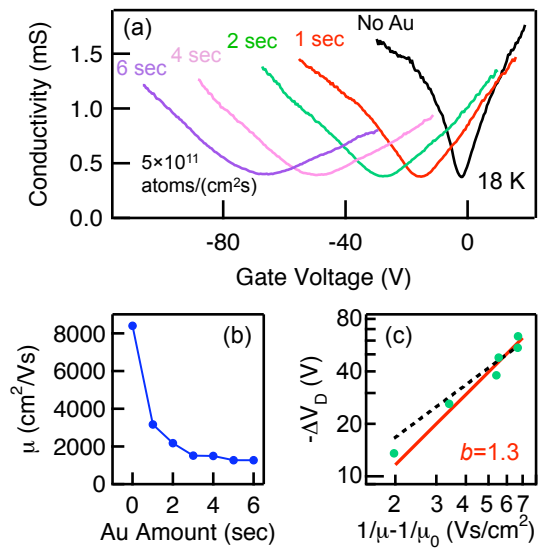

Figure 1

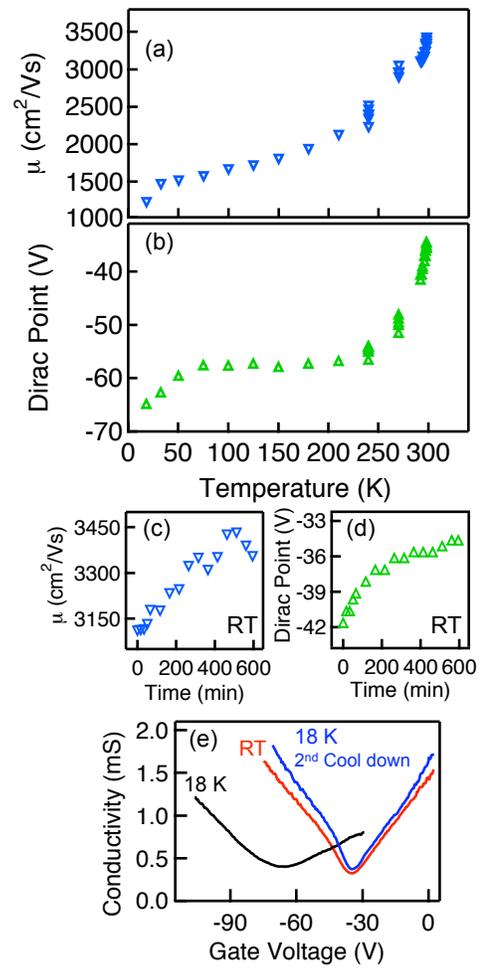

Figure 2

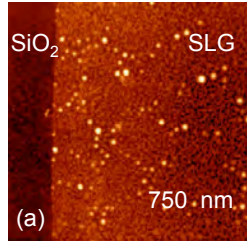
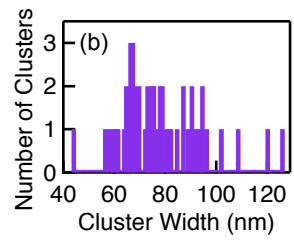

Figure 3